\documentclass[%
twocolumn,
reprint,
nosuperscriptaddress,
nofootinbib,
showpacs,
notitlepage,
showkeys,
nobibnotes,
 amsmath,amssymb,
 aps, prd,
floatfix,
]{revtex4-1}

\usepackage{graphicx}
\usepackage{dcolumn}
\usepackage{bm}

\usepackage[unicode,pdfauthor={S. Shashank, C. Bambi}, pdftitle={Constraining KRZ III: limits from GW data}]{hyperref}
\hypersetup{colorlinks,urlcolor=blue,citecolor=blue,linkcolor=blue}
\usepackage{amssymb}
\usepackage{amsmath}
\usepackage{amsfonts}
\usepackage{envmath}
\usepackage{float}
\usepackage{verbatim}

\newcommand{\be}{\begin{eqnarray}}
\newcommand{\ee}{\end{eqnarray}}

\begin{document}

\preprint{APS/123-QED}

\title{Constraining the Konoplya-Rezzolla-Zhidenko deformation parameters III:\\limits from stellar-mass black holes using gravitational-wave observations}

\newcommand{\Fudan}{Center for Field Theory and Particle Physics and Department of Physics, Fudan University, 200438 Shanghai, China}

\author{Swarnim Shashank}
\affiliation{\Fudan}

\author{Cosimo Bambi}
\email[Corresponding author: ]{bambi@fudan.edu.cn}
\affiliation{\Fudan}

\begin{abstract}
Gravitational-wave observations of binary black holes provide a suitable arena to test the fundamental nature of gravity in the strong-field regime. Using the data of the inspiral of 29 events detected by the LIGO-Virgo observatories, we perform a theory-agnostic test of the Kerr hypothesis. We compute the leading-order deviation to the gravitational waves emitted in the frequency domain and provide constraints on two deformation parameters ($\delta_1$ and $\delta_2$) belonging to a general class of axisymmetric non-Kerr black hole spacetimes proposed by Konoplya, Rezzolla \& Zhidenko. Our study shows that all the analyzed events are consistent with the Kerr hypothesis. The LIGO-Virgo data provide stronger constraints on $\delta_1$ and $\delta_2$ than those obtained in our previous studies with X-ray data (Papers~I and II), while, on the other hand, they cannot constrain the other deformation parameters of the Konoplya-Rezzolla-Zhidenko metric ($\delta_3$, $\delta_4$, $\delta_5$, and $\delta_6$).
\end{abstract}

\maketitle

\section{Introduction}\label{sec:intro}

In February 2016, the Laser Interferometer Gravitational-Wave Observatory (LIGO) and Virgo collaboration announced the first direct detection of gravitational waves (GWs) from the merger of two black holes (BHs)~\cite{first_gw_bbh_2016}. GW astronomy provides us a new window of observing astrophysical events in the strong gravity regime, allowing us to test general relativity (GR) for these events~\cite{Yunes_Yagi_Pretorius_2016}. To date, the LIGO/Virgo collaboration (LVC) has finished their third observing run (O3a) and reported a total of 46 binary BH merger events~\cite{GWTC1,GWTC2}.

To test GR as a theory of gravity, one of the approaches is to use a parametrized metric and measure possible deviations from the solution of Einstein equations such as the Kerr solution \cite{Johannsen:2013szh,Konoplya_Rezzolla_Zhidenko_2016,Ghasemi-Nodehi:2016wao,Mazza:2021rgq}. In the past few years, there have been a number of studies testing the Kerr hypothesis~\cite{Bambi:2015kza} by using X-ray data from \textsl{NuSTAR}, \textsl{RXTE}, \textsl{Suzaku}, and \textsl{XMM-Newton}~\cite{Cao:2017kdq,Tripathi:2018lhx,Tripathi:2020qco,Tripathi:2020dni,Tripathi:2020yts,Zhang:2021ymo}, GW data from LIGO and Virgo~\cite{Cardenas-Avendano_Nampalliwar_Yunes_2020,Carson:2020iik,Psaltis:2020ctj}, and radio data from the Event Horizon Telescope (EHT) experiment~\cite{Bambi:2019tjh,EventHorizonTelescope:2020qrl,Volkel:2020xlc,Psaltis:2020ctj}.

In this work, we follow-up on the study of the Konoplya-Rezzolla-Zhidenko (KRZ) parameterized BH metric~\cite{Konoplya_Rezzolla_Zhidenko_2016}, which was done earlier using X-ray reflection spectroscopy (XRS)~\cite{Bambi:2020jpe} for the supermassive BH in the galaxy MCG--06--30--15 \cite{Abdikamalov:2021zwv} (hereafter Paper~I) and the stellar-mass BH in the binary system EXO~1846--031 \cite{Yu:2021xen} (hereafter Paper~II). In the present study we still constrain the parameters of the KRZ metric for stellar-mass BHs, but we use GW data. For our analysis, we use binary black hole (BBH) data from LVC published in their first (GWTC-1 \cite{GWTC1}) and second (GWTC-2 \cite{GWTC2}) catalogs, which have events observed up to 1 October 2019 \cite{ligo_posterior_gwtc1, ligo_posterior_gwtc2}.

The manuscript is organised as follows. In Sec.~\ref{sec:krz_metric}, we revisit the KRZ metric along with a discussion of equatorial geodesics in the KRZ spacetime. Sec.~\ref{sec:gw_constraint} shows our methodology for testing the Kerr nature of BHs using the KRZ deformation parameters. Sec.~\ref{sec:results} shows our results. In Sec.~\ref{sec:conclusion}, we compare our results with the constraints inferred from X-ray in Paper~I and Paper~II and we end with some concluding remarks. Throughout the article, we use units in which $c = G_{\rm N} = 1$ and the metric signature $(- + + +)$.

\section{The KRZ metric}\label{sec:krz_metric}

\subsection{Revisiting the KRZ metric} \label{subsec:krz_revisit}

We adopt Boyer-Lindquist coordinates and we write the KRZ metric in the following form \cite{Nampalliwar:2019iti, Konoplya_Rezzolla_Zhidenko_2016}:
\begin{widetext}
\be
d s^{2} = -\frac{N^{2}-W^{2} \sin ^{2} \theta}{K^{2}} d t^{2} + \frac{\Sigma B^{2}}{N^{2}} d r^{2}+\Sigma r^{2} d \theta^{2} - 2 W r \sin ^{2} \theta d t d \phi + K^{2} r^{2} \sin ^{2} \theta d \phi^{2} \, ,
\label{eq:krz_metric}
\ee
where
\be
N^{2} &=& \left(1-\frac{r_{0}}{r}\right)\left(1-\frac{\epsilon_{0} r_{0}}{r}+\left(k_{00}-\epsilon_{0}\right) \frac{r_{0}^{2}}{r^{2}}+\frac{\delta_{1} r_{0}^{3}}{r^{3}}\right) + \left(\frac{a_{20} r_{0}^{3}}{r^{3}}+\frac{a_{21} r_{0}^{4}}{r^{4}}+\frac{k_{21} r_{0}^{3}}{r^{3}\left(1+\frac{k_{22}\left(1-\frac{r_0}{r}\right)}{1+k_{23}\left(1-\frac{r_0}{r}\right)}\right)}\right) \cos ^{2} \theta \, , \\
B &=& 1+\frac{\delta_{4} r_{0}^{2}}{r^{2}}+\frac{\delta_{5} r_{0}^{2}}{r^{2}} \cos ^{2} \theta \, , \\
\Sigma &=&1+\frac{a_{*}^{2} M^2}{r^{2}} \cos ^{2} \theta \, , \\
W &=& \frac{1}{\Sigma}\left(\frac{w_{00} r_{0}^{2}}{r^{2}}+\frac{\delta_{2} r_{0}^{3}}{r^{3}}+\frac{\delta_{3} r_{0}^{3}}{r^{3}} \cos ^{2} \theta\right) \, , \\
K^{2} &=& 1+\frac{a_{*} M W}{r} + \frac{1}{\Sigma}\left(\frac{k_{00} r_{0}^{2}}{r^{2}}+\frac{k_{21} r_{0}^{3}}{r^{3}\left(1+\frac{k_{n}\left(1-\frac{r_0}{r}\right)}{1+k_{23}\left(1-\frac{r_0}{r}\right)}\right)} \cos ^{2} \theta\right) \, ,
\ee
and
\be
&& r_{0}=M \left( 1+\sqrt{1-a_{*}^{2}} \right) \, , \qquad
\epsilon_{0}=\frac{2M-r_{0}}{r_{0}} \, , \qquad
a_{20}=\frac{2 a_{*}^{2}M^3}{r_{0}^{3}} \, , \qquad 
a_{21}=-\frac{a_{*}^{4}M^4}{r_{0}^{4}}+\delta_{6} \, , \nonumber\\
&& k_{00}=k_{22}=k_{23}=\frac{a_{*}^{2}M^2}{r_{0}^{2}} \, , \qquad
k_{21}=\frac{a_{*}^{4}M^4}{r_{0}^{4}}-\frac{2a_{*}^{2}M^3}{r_{0}^{3}}-\delta_{6} \, , \qquad
w_{00}=\frac{2 a_{*}M^2}{r_{0}^{2}} \, ,
\ee
\end{widetext}
where $M$ is the BH mass, $a_{*}=J/M^2$ is the dimensionless BH spin parameter, and $r_0$ turns out to be the radius of the BH event horizon.

The metric quantifies deviations from the Kerr metric via six deformation parameters viz. $\{ \delta_i \} \text{ } (\text{where } i = 1,2,...,6)$. The KRZ metric reduces to the Kerr solution when all the parameters are identically set to zero\footnote{The metric does not reduce to the Kerr solution in the expression provided in Ref. \cite{Konoplya_Rezzolla_Zhidenko_2016}; the modified expression can be found in Ref.~\cite{Nampalliwar:2019iti}.}. The deformation parameters have their physical interpretations which could help mapping the metric to other theories of gravity; they are listed as follows \cite{Konoplya_Rezzolla_Zhidenko_2016, Nampalliwar:2019iti}:
$$
\begin{aligned}
\delta_{1} & \rightarrow \text { deformations on } g_{t t} \\
\delta_{2}, \delta_{3} & \rightarrow \text { rotational deformations} \\
\delta_{4}, \delta_{5} & \rightarrow \text { deformations on } g_{r r} \\
\delta_{6} & \rightarrow \text { deformations on the event horizon.}
\end{aligned}
$$
Bounds can be set on the deformation parameters $\{ \delta_i \}$ to avoid certain pathologies in the spacetime outside of the horizon. These can be avoided by following certain conditions: the metric determinant must be always negative, the metric coefficient $g_{\phi \phi}$ must be greater than zero, and $N^2$ must be non-vanishing \cite{Nampalliwar:2019iti}. Implying these conditions gives the following bounds on the KRZ deformation parameters:
\begin{eqnarray}
    \begin{aligned}\label{eq-bounds}
        \delta_1 &> \frac{4 M r_{0}-3 r_{0}^{2}-a_{*}^{2} M^2}{r_{0}^{2}} \\
        \delta_{2}, \delta_{3} & \begin{cases} > \\ < \end{cases} -\frac{4}{a_{*}^3}\left(1-\sqrt{1-a_{*}^{2}}\right) \begin{aligned} &\text { if } a_{*}>0 \\ &\text { if } a_{*}<0 \end{aligned} \\
        \delta_{4}, \delta_{5} &> -1 \\
        \delta_{6} &< \frac{r_{0}^{2}}{M^2 \left(4-a_{*}^{2}\right)} \, .
    \end{aligned}
\end{eqnarray}

Currently, for both X-ray and GW data, only a single deformation parameter can be constrained at a time and the other parameters are set to zero during the analysis. This is done to reduce the complexity of the analysis. However, it could be expected that multiple non-vanishing deformation parameters are necessary to recover non-Kerr BH solutions in other theories of gravity.

\subsection{The case for equatorial geodesics}\label{subsec:equa_geodesic}

To study the GWs from a spacetime parametrized by the KRZ metric, we consider only the inspiral phase of the binary system, which can be studied using the post-Newtonian (PN) formalism \cite{Blanchet_2014}. For inspirals, we consider equatorial geodesics; i.e., $\theta = \pi/2$ and $\dot{\theta} = 0$. As a result of this assumption, the only deformation parameters that remain are $\delta_1$, $\delta_2$, and $\delta_4$ (the terms with $\delta_3$, $\delta_5$, and $\delta_6$ are all proportional to $\cos ^{2} \theta$ and vanish for equatorial orbits). Also, we are working in the leading order approximation for the deformation parameters that enter at a smaller PN order than spin. The parameter $\delta_1$ enters at 2PN and $\delta_2$ enters at 2.5PN (see Sec.~\ref{sec:gw_constraint}, Eq.~\ref{eq:mod_kep_delta1} and Eq.~\ref{eq:mod_kep_delta2}). The spin-contributions only enter at 3.5PN level or higher. Hence, we set $a_* = 0$, which implies $r_0=2M$. The metric becomes
\be
    d s^{2}&=&-\frac{N^{2}-W^{2}}{K^{2}} ~d t^{2}-2 W r ~d t d \phi \nonumber\\
    &&+K^{2} r^{2} ~d \phi^{2} + \Sigma r^2 d\theta^{2} + \frac{\Sigma B^{2}}{N^{2}} ~  d r^{2} \, , 
\label{eq:krz_metric_equatorial}
\ee
where
\be
    N^2 &=& \left( 1 - \frac{2M}{r} \right)~\left[ 1 + \frac{8 M^3 \delta_1}{r^3} \right] \, , \\
    B &=& 1 + \frac{4 M^2 \delta_4}{r^2} \, , \\
    \Sigma &=& 1, \quad K^2 = 1 \, , \\
    W &=& \frac{8 M^3 \delta_2}{r^3} \, .
\label{eq:coeff}
\ee
In such a limit, the bounds in Eq.~\ref{eq-bounds} become $\delta_1, \delta_4 > -1$ and there is no bound for $\delta_2$.

The specific energy and the specific angular momentum of a test-particle are
\be\label{eq:specific_energy}
    E &=& -g_{t \mu} u^{\mu} =\left(\frac{N^{2}-W^{2}}{K^{2}}\right) \dot{t} + 2 W r \dot{\phi} \, , \\
    L &=& g_{\phi \mu} u^{\mu}=-2 W r \dot{t}+K^{2} r^{2} \dot{\phi} \, , \label{eq:specific_ang_mom} 
\ee
and are constants of motion because the spacetime is stationary and axisymmetric.
From Eqs.~\ref{eq:specific_energy} and \ref{eq:specific_ang_mom}, we can write $\dot{t}$ and $\dot{\phi}$
\be
    \dot{t} &=& -\frac{2 L W - E K^{2} r}{r (N^{2}+3 W^2)} \, , \\
    \dot{\phi} &=& -\frac{-2 r E K^{2} W - (N^{2} - W^2)~L}{r^2 K^{2} (N^{2} + 3 W^2 )} \, .
\ee

\section{Gravitational wave constraints}\label{sec:gw_constraint}

Within the approximations discussed in the previous section, only the deformation parameters $\delta_1$ and $\delta_2$ remain when considering equatorial geodesics. For $\delta_4$, we obtain the same Kepler's law as that for Kerr: $\delta_4$ does not enter the geodesic motion in our case (see later). In this paper, when we want to constrain one of the deformation parameters we assume that all other deformation parameters vanish.

In the PN formalism, a two-body problem can be studied as an effective one body problem where a test particle of reduced mass $\mu = m_1 m_2 / m$ ($m = m_1 + m_2$, and $m_1$ and $m_2$ are the masses of the two bodies), follows the geodesic motion around an object of mass $m$. The four-momentum of the test particle is contracted with the KRZ metric to give us the effective Hamiltonian, which is responsible for the conservative sector of the orbital motion \cite{Buonanno:1998gg, Hinderer:2017jcs}. This effective problem then can be mapped back to the two-body problem to give us the GWs emitted by such a parametrically deformed BBH system.
Here we follow the analysis proposed in Ref.~\cite{Cardenas-Avendano_Nampalliwar_Yunes_2020}. We note that the $\delta_1$ deformation parameter results to be the same as the bumpy parameter $a_1$ of the RZ metric \cite{Rezzolla_Zhidenko_2014} in Ref.~\cite{Cardenas-Avendano_Nampalliwar_Yunes_2020}. Nevertheless, in this work we extend that analysis for the events in the GWTC-2 catalog.

\subsection{Deformation parameter $\delta_1$}\label{subsec:delta1}

First, we consider the possibility of a non-vanishing deformation parameter $\delta_1$. From the normalization condition of the four-velocity, $u^{\mu} u_{\mu} = -1$, we have ($\dot{\theta} = 0$)~\cite{Bambi:2017khi}
\be
g_{rr} \dot{r}^2 = -1 - g_{tt} \dot{t}^2 - g_{\phi\phi} \dot{\phi}^2 \equiv V_{\mathrm{eff}} \, ,
\ee
where we have introduced the effective potential $V_{\mathrm{eff}}$, which can be written in terms of $E$ and $L$ as
\be
\label{eq:Veff_full_delta1}
    V_{\mathrm{eff}} &=& -1 + E^2 + \frac{2 M}{r} + \frac{L^2 (2 M-r)}{r^3} + \frac{8 \delta_1 M^3 (2 M-r)}{r^4} \nonumber\\ 
    && + \frac{8 \delta_1 L^2 M^3 (2 M-r)}{r^6} +\mathcal{O}[{\delta_1}^{2}] \, .
\ee

The effective potential can be written as the standard effective potential of the Kerr metric plus a small deformation.
So we expand this effective potential about the small deformation away from Kerr
\be
\label{eq:eff_pot}
V_{\mathrm{eff}}=V_{\mathrm{eff}}^{\mathrm{GR}}+V_{\mathrm{eff}}^{\mathrm{KRZ}}+\mathcal{O}[{\delta_1}^{2}] \, ,
\ee
where
\be
V_{\mathrm{eff}}^{\mathrm{GR}} &=& E^2 + \frac{(L^2+r^{2})(2 M-r)}{r^{3}} \, , \\
    V_{\mathrm{eff}}^{\mathrm{KRZ}} &=& \frac{8 M^3 (L^2+r^2) (2 M-r)}{r^6} ~ \delta_1 \, .
\ee

The energy and angular momentum for circular orbits can be found by the condition $V_{\mathrm{eff}} = dV_{\mathrm{eff}}/dr = 0$~\cite{Bambi:2017khi} and they can be rewritten as the standard GR term plus a small contribution that depends on the deformation parameter $\delta_1$
\begin{eqnarray}
	\begin{aligned}\label{eq:E_L_expansion}
		E &=& E^{\mathrm{GR}} + \delta E \, , \\
    		L &=& L^{\mathrm{GR}} + \delta L \, ,
	\end{aligned}
\end{eqnarray}
where
\be
    E^{\mathrm{GR}} &=& \sqrt{\frac{4 M^{2}-4 M r+r^{2}}{(r-3 M)r}} \, , \\
    L^{\mathrm{GR}} &=& \sqrt{\frac{M r^{2}}{r-3 M}} \, , 
\ee
and
\be
    \delta E &=& -\frac{2 M^3 (r-2 M)}{r^{5/2} (r-3 M)^{3/2}} \delta_1 + \mathcal{O} [{\delta_1}^{2}] \, , \\
    \delta L &=& -\frac{6 M^{5/2} (r-2 M)^2}{r^2 (r-3 M)^{3/2}} \delta_1 + \mathcal{O} [{\delta_1}^{2}] \, .
\ee

In the far-field limit, $L = r^2 \dot{\phi} \rightarrow r^2 \Omega$, where $\Omega = d\phi/dt$ is the angular velocity of the body as measured by a distant observer, and we find
\be \label{eq:mod_kep_delta1}
    \Omega^2 = \frac{M}{r^3} \left[ 1 + \frac{3 M}{r} + \frac{9 M^2}{r^2} - \frac{12 M^2}{r^2}\delta_1 + \mathcal{O} \left({\delta_1}^{2}, \frac{M^3}{r^3} \right) \right] \, . \nonumber\\
\ee
Here the power of $M$ in the $\delta_1$ term represents the PN order \cite{Carson:2020iik}, so we can say that $\delta_1$ enters at 2PN.

In order to map our results back to the two-body problem, we use the total energy and the effective energy \cite{Damour:2000kk, Buonanno:1998gg}. For circular orbits, the total energy ($E_T$) of the system can be written in terms of the effective energy, which is the energy of one body in the rest frame of the other body \cite{Damour:2000kk, Buonanno:1998gg}:
\be
E_\mathrm{T} = m + E_\mathrm{b} = m [ 1 + 2 \eta (E_\mathrm{eff} - 1) ] \, ,    
\ee
where 
\be
E_\mathrm{eff} = g_{tt} \left( 1 + \frac{L^2}{r^2} \right)^{1/2} \, ,
\ee
$E_\mathrm{b}$ is the binding energy, and $\eta = \mu/m$ is the symmetric mass ratio. The rest-mass energy $m$ and the binding energy $E_\mathrm{b}$ are separated in order to write the latter as its GR term plus a correction
\be
    E_\mathrm{b} = E_\mathrm{b}^\mathrm{GR} - \frac{\eta m^2}{2r} \left[ 4 ~\delta_1 \left( \frac{m}{r} \right)^2 + \mathcal{O} \left( {\delta_1}^{2}, \frac{m^3}{r^3} \right) \right] \, . \nonumber\\
\ee

Since the angular frequencies of the effective one-body problem and of the two-body problem are the same, it is convenient to write the binding energy $E_\mathrm{b}$ in terms of the orbital frequency $\nu = \Omega/2\pi$,
\be
    \frac{E_\mathrm{b}(\nu)}{\mu} = \frac{E_\mathrm{b}^\mathrm{GR}(\nu)}{\mu} - 4 \delta_1(2 \pi m \nu)^{2}+\mathcal{O}\left[{\delta_1}^{2},(2 \pi m \nu)^{8 / 3}\right] \, . \nonumber\\
\ee
The orbital phase is given by \cite{Cardenas-Avendano_Nampalliwar_Yunes_2020, Blanchet_2014}
\be
    \phi(\nu) = \int^{\nu} \Omega ~dt 
    = \int^{\nu} \frac{1}{\dot{E}} \left(\frac{dE}{d \Omega}\right)  \Omega ~ d \Omega \, ,
\ee
where $\dot{E}$ represents the rate of change of binding energy caused by the emission of GWs. The latter depends on both the conservative sector (binding energy) and the dissipative sector (energy loss rate). However, X-ray constraints are only sensitive to the conservative sector and we are comparing our results to those in Paper~I and Paper~II, so we consider only modifications on the conservative dynamics and we assume dissipative dynamics to be the same as GR (for a more detailed discussion, see Ref.~\cite{Cardenas-Avendano_Nampalliwar_Yunes_2020}). Hence, we only need to use the quadrupole formula to the leading PN order (0PN) for the change in binding energy \cite{Cardenas-Avendano_Nampalliwar_Yunes_2020, Blanchet_2014}; i.e., 
\be
\dot{E}_\mathrm{GR}^\mathrm{0PN} = -(32 / 5) \eta^{2} m^{2} r^{4} \Omega^{6} \, .
\ee

Thus we obtain the expression for the orbital phase evolution
\be
    \phi(\nu) = \phi_{\mathrm{GR}}(\nu)-\frac{25}{4 \eta}(2 \pi m \nu)^{-1 / 3} \delta_1+\mathcal{O}\left[{\delta_1}^{2},(2 \pi m \nu)^{0}\right] \, , \nonumber\\
\ee
where
\be
\phi_{\mathrm{GR}}^\mathrm{0PN}(\nu)=-\frac{1}{32 \eta}(2 \pi m \nu)^{-5/3} \, .
\ee

The Fourier transform of $\phi$ in the stationary phase approximation is $\Psi_{\mathrm{GW}}(f)=2 \phi (t_0) - 2 \pi f t_0$, where $t_0$ is the stationary time such that $\nu (t_0) = f/2$ and $f$ is the Fourier frequency \cite{Cardenas-Avendano_Nampalliwar_Yunes_2020}. At the leading order in the deformation parameter, we have
\be\label{eq:GW_phase_delta1}
    \Psi_{\mathrm{GW}}(f)=\Psi_{\mathrm{GW}}^{\mathrm{GR}}(f)-\frac{75}{8} u^{-1/3} \eta^{-4/5} \delta_1 + \mathcal{O}[{\delta_1}^{2}, u^0] \, , \nonumber\\
\ee
where
\be
\Psi_{\mathrm{GW}}^{\mathrm{GR}, 0 \mathrm{PN}}(f) = - \frac{3 u^{-5/3}}{128} \, 
\ee
and $u = \eta^{3/5} \pi m f$.

We compare Eq.~\ref{eq:GW_phase_delta1} to the parameterized post-Einsteinian (ppE) framework \cite{Yunes_Pretorius_2009}, where
\be
\label{eq:ppE}
    \Psi_{\mathrm{GW}} = \Psi_{\mathrm{GW}}^{\mathrm{GR}} + \beta u^{b}
\ee
and $b = -1/3$ at 2PN. We find
\be
\label{eq:beta_ppe}
    \beta = -\frac{75}{8} \eta^{-4/5} \delta_1 \, .
\ee

The ppE parameterization used by LVC is \cite{Yunes_Yagi_Pretorius_2016, lalsuite}
\be \label{eq:beta_ligo}
    \beta = \frac{3}{128} \varphi_{4} \delta \varphi_{4} \eta^{-4/5} \, .
\ee
$\varphi_{4}$ is the PN phase at 2PN and has the form \cite{Khan_2016_posterior}
\be
    \varphi_{4} = \frac{15293365}{508032}+\frac{27145 \eta}{504}+\frac{3085 \eta^{2}}{72} \, .
\ee
$\delta \varphi_4$ is the deviation from GR given as correction to the non-spinning portion of the PN phase \cite{ligo_posterior_gwtc1, ligo_posterior_gwtc2}
\be
    \varphi_i \rightarrow (1 + \delta \varphi_i) \varphi_i \, .
\ee
Comparing Eqs. \ref{eq:beta_ppe} and \ref{eq:beta_ligo}, we find
\be\label{eq:d1-p4}
   \delta_1 = -\frac{1}{400} \varphi_{4} \delta \varphi_{4} \, .
\ee

\subsection{Deformation parameter $\delta_2$}\label{subsec:delta2}

For the deformation parameter $\delta_2$, we follow the method discussed in Subsec.~\ref{subsec:delta1}. We write the effective potential assuming that all the other deformation parameters vanish and we find
\be\label{eq:Veff_full_delta2}
    V_{\mathrm{eff}} &=& -1 + E^2 + \frac{2 M}{r} + \frac{L^2 (2 M-r)}{r^3} \nonumber\\
    && + \frac{32 E L M^3 \delta_2}{r^4} + \mathcal{O} [{\delta_2}^{2}] \, .
\ee
As done in Eq.~\ref{eq:eff_pot}, we separate the GR and non-GR part. The non-GR part without $\mathcal{O}[{\delta_2}^2]$ corrections is now
\be
    V_{\mathrm{eff}}^{\mathrm{KRZ}} = \frac{32 E L M^3}{r^4} ~ \delta_2 \, .
\ee
The corrections to the energy and angular momentum of circular orbits are now
\be
    \delta E &=& -\frac{16 M^{7/2} (M-r)}{r^3 (r-3 M)^{3/2}} \delta_2 + \mathcal{O} [{\delta_2}^{2}] \, , \\
    \delta L &=& -\frac{32 M^3 (2 M-r)}{(r (r-3 M))^{3/2}} \delta_2 + \mathcal{O} [{\delta_2}^{2}] \, .
\ee

The modified Kepler's law reads
\be
\label{eq:mod_kep_delta2}
    \Omega^2 = \frac{M}{r^3}\left[ 1 + \frac{3 M}{r} + \frac{9 M^2}{r^2} + \frac{64 M^{5/2}}{r^{5/2}}\delta_2 + \mathcal{O} \left({\delta_2}^{2}, \frac{M^3}{r^3} \right) \right] \, ,\nonumber\\
\ee
where, again, the exponent of $M$ in the $\delta_2$ term represents the PN order \cite{Carson:2020iik}. 
The binding energy is now
\be
    E_\mathrm{b} = E_\mathrm{b}^\mathrm{GR} + \frac{\eta m^2}{2r} \left[ 32 ~\delta_2 \left( \frac{m}{r} \right)^{2.5} + \mathcal{O} \left( {\delta_2}^{2}, \frac{m^3}{r^3} \right) \right] \, . \nonumber\\
\ee

As done in the previous section, we write the orbital phase
\be
    \phi(\nu) &=& \phi_{\mathrm{GR}}(\nu) \nonumber\\ && - \left(\frac{85 \log (\nu)}{6 \eta } + \frac{85 \log (2)}{6 \eta } + \frac{85 \log (\pi )}{6 \eta }\right) \delta_2 \, , \nonumber\\
\ee
and then its Fourier transform
\be
\label{eq:GW_phase_delta2}
    \Psi_{\mathrm{GW}}(f) = \Psi_{\mathrm{GW}}^{\mathrm{GR}}(f) - \frac{85}{3 \eta} [1 + \log(u)] \delta_2 \, .
\ee

In the ppE framework, $b = 0$ for the logarithmic term at 2.5PN \cite{Yunes_Yagi_Pretorius_2016,lalsuite} and we find
\be
\beta = - \frac{85}{3 \eta} [1 + \log(u)] \delta_2 \, .
\ee
In the ppE parametization used by LVC
\be
 \beta = \frac{3}{128} \varphi_5 \delta \varphi_5 \eta^{-1} \, ,
\ee
where $\varphi_5$ is \cite{Khan_2016_posterior}
\be \label{eq:phi5l}
    \varphi_{5} = [1+\log(u)] \left[\frac{38645 \pi}{756}-\frac{65 \pi \eta}{9} \right] \, ,
\ee
and $\delta \varphi_5$ is the non-GR PN phase. The mapping on $\delta_2$ is
\be\label{eq:d2-p5}
    \delta_2 = -\left(\frac{7729 \pi}{182784}-\frac{13 \pi \eta}{2176} \right) \delta \varphi_5 \, .
\ee

\subsection{Deformation parameter $\delta_4$}\label{subsec:delta4}

Setting $\delta_1=\delta_2=0$, the effective potential for $\delta_4$ is
\be\label{eq:Veff_full_delta4}
    V_{\mathrm{eff}} &=& -1 + E^2 + \frac{2 M}{r} + \frac{L^2 (2 M-r)}{r^3} \nonumber\\
    && -\frac{8 L^2 M^2 (2 M-r) \delta_4}{r^5} -\frac{8 M^2 (2 M-r) \delta_4}{r^3} \nonumber\\
    && -\frac{8 E^2 M^2 \delta_4}{r^2} + \mathcal{O} [{\delta_4}^{2}]
\ee
and the leading-order non-GR correction is
\be
    V_{\mathrm{eff}}^{\mathrm{KRZ}} = -\frac{8 M^2 \left(\left(L^2+r^2\right) (2 M-r)+E^2 r^3\right)}{r^5}~\delta_4 \, .
    \nonumber\\
\ee
However, the non-GR corrections to the energy and angular momentum of circular orbits vanish: $\delta E = \delta L = 0$. The deformation parameter $\delta_4$ does not alter the motion of equatorial geodesics and therefore it cannot be constrained within our framework.

\section{Results} \label{sec:results}

We fit the publicly available posterior samples released by the LVC~\cite{ligo_scientific_collaboration_and_virgo_2021_5172704} to obtain the constraints on the deformation parameters $\delta_1$ and $\delta_2$. We use the publicly available Markov Chain Monte Carlo (MCMC) sample for the best-fit model inferred by two independent analysis, ``Tests of General Relativity with Binary Black Holes from the second LIGO-Virgo Gravitational-Wave Transient Catalog - Full Posterior Sample Data Release"\footnote{\url{https://zenodo.org/record/5172704}}. For the names of the events, we follow the convention from the scripts used by LVC in Ref.~\cite{ligo_scientific_collaboration_and_virgo_2021_5172704}.

From the posterior samples, we use the individual red-shifted (detector frame) mass of the binaries and non-GR parameters $\delta \varphi_4$ and $\delta \varphi_5$, and we calculate the samples for $\delta_1$ and $\delta_2$. Our results are shown in Tab.~\ref{tab:delta1} for $\delta_1$ and in Tab.~\ref{tab:delta2} for $\delta_2$, where we report the uncertainties at the 90\% confidence limit. All the events are BBH merger sources. One of the events is GW190814A \cite{GW190814:2020zkf} which has components of mass $23.2~M_{\odot}$ and $2.59~M_{\odot}$ \cite{GWTC2}, where the second source may either be the lightest BH or the heaviest neutron star observed. However, in a recent study, it was stated that the secondary source in GW190814A was most likely a BH~\cite{Nathanail:2021tay}. Hence, GW190814A is included in our analysis but our constraint only holds if the secondary source is a BH.

We report constraints obtained from the two waveform models, i.e., IMRPhenomPv2 and SEOBNRv4P, used by the LVC~\cite{ligo_posterior_gwtc2}. The phenomenological model IMRPhenomPv2~\cite{Hannam:2013oca} describes GWs from precessing BBH systems, is calibrated to numerical relativity, and uses an effective single-spin description to model effects from spin-precession. On the other hand, SEOBNRv4P~\cite{Bohe:2016gbl} follows the effective-one-body formalism, and calibrated to Numerical Relativity with a generic two-spin treatment of the precession dynamics. We note that for some events in the GWTC-2 catalog, the IMPRPhenomPv2 or SEOBNRv4P data are not available~\cite{ligo_posterior_gwtc2}; in those cases, only one constraint is reported.

\begin{table}
    \begin{ruledtabular}
    \begin{tabular}{ccc}
        Event & $\delta_1$ (IMRPhenomPv2) & $\delta_1$ (SEOBNRv4P) \\
        \hline
        GW150914 & $0.22_{-0.19}^{+0.18}$ & $0.19_{-0.15}^{+0.17}$ \\
		GW151226 & $-0.01_{-0.18}^{+0.15}$ & $-0.03_{-0.20}^{+0.18}$ \\
		GW170104 & $-0.43_{-0.53}^{+0.41}$ & $-0.25_{-0.45}^{+0.33}$ \\
		GW170608 & $0.03_{-0.12}^{+0.12}$ & $0.05_{-0.15}^{+0.16}$ \\
		GW170814 & $0.05_{-0.20}^{+0.19}$ & $-0.02_{-0.22}^{+0.20}$ \\
		
		GW190408A & $-0.17 ^{+0.32} _{-0.31}$ & $-0.11 ^{+0.39} _{-0.31}$ \\
		GW190412A & $0.06 ^{+0.14} _{-0.12}$ & $0.07 ^{+0.13} _{-0.12}$ \\
		GW190421A & $7.44 ^{+3.53} _{-16.26}$ & -- \\
		GW190503A & $-0.42 ^{+1.13} _{-6.82}$ & $-0.23 ^{+0.63} _{-6.46}$ \\
		GW190512A & $-0.74 ^{+1.07} _{-0.31}$ & $0.22 ^{+0.23} _{-1.27}$ \\
		GW190513A & $-0.51 ^{+0.57} _{-0.93}$ & $-0.70 ^{+0.76} _{-0.66}$ \\
		GW190517A & $0.39 ^{+1.55} _{-1.21}$ & $0.50 ^{+0.75} _{-0.58}$ \\
		GW190519A & $-1.19 ^{+31.07} _{-29.12}$ & -- \\
		GW190521A & $-0.03 ^{+3.84} _{-3.90}$ &  --\\
		GW190521B & $0.08 ^{+0.21} _{-0.25}$ & $0.01 ^{+0.20} _{-0.17}$ \\
		GW190602A & $-0.17 ^{+2.13} _{-1.85}$ & $-0.41 ^{+2.27} _{-1.64}$ \\
		GW190630A & $-0.20 ^{+0.28} _{-0.27}$ & $-0.19 ^{+0.25} _{-0.26}$ \\
		GW190706A & $0.10 ^{+1.93} _{-2.08}$ &  --\\
		GW190707A & $-0.07 ^{+0.16} _{-0.16}$ & $-0.08 ^{+0.17} _{-0.22}$ \\
		GW190708A & $-0.05 ^{+0.17} _{-0.19}$ & $-0.06 ^{+0.16} _{-0.18}$ \\
		GW190720A & $-0.08 ^{+0.17} _{-0.21}$ & $-0.08 ^{+0.20} _{-0.20}$ \\
		GW190727A & $3.56 ^{+27.29} _{-34.74}$ & $1.36 ^{+0.70} _{-3.19}$ \\
		GW190728A & $-0.14 ^{+0.21} _{-0.27}$ &  --\\
		GW190814A & $-0.01 ^{+0.60} _{-0.08}$ &  --\\
		GW190828A & $0.09 ^{+0.27} _{-0.31}$ & $0.11 ^{+0.23} _{-0.23}$ \\
		GW190828B & $-0.31 ^{+0.32} _{-0.34}$ & $-0.35 ^{+0.35} _{-0.52}$ \\
		GW190910A & $0.00 ^{+2.36} _{-0.52}$ & $0.03 ^{+0.31} _{-0.33}$ \\
		GW190915A & -- & $-0.08 ^{+2.18} _{-2.05}$ \\
		GW190924A & $-0.09 ^{+0.24} _{-0.37}$ & $-0.18 ^{+0.40} _{-0.60}$ \\
    \end{tabular}
    \end{ruledtabular}
    \caption{Constraints on $\delta_1$ for the BBH events in GWTC-1 and GWTC-2 with the IMRPhenomPv2 and SEOBNRv4P waveform models. The reported uncertainties correspond to the 90\% confidence limit. -- means that the data are not available. See the text for more details.}
    \label{tab:delta1}
\end{table}

\begin{table}
    \begin{ruledtabular}
    \begin{tabular}{ccc}
        Event & $\delta_2$ (IMRPhenomPv2) & $\delta_2$ (SEOBNRv4P) \\
        \hline
        GW150914 & $-0.09_{-0.07}^{+0.07}$ & $-0.08_{-0.06}^{+0.06}$ \\
		GW151226 & $0.00_{-0.06}^{+0.09}$ & $0.00_{-0.09}^{+0.10}$ \\
		GW170104 & $0.18_{-0.16}^{+0.21}$ & $0.12_{-0.14}^{+0.20}$ \\
		GW170608 & $-0.01_{-0.05}^{+0.05}$ & $-0.03_{-0.08}^{+0.07}$ \\
		GW170814 & $-0.01_{-0.09}^{+0.08}$ & $0.02_{-0.08}^{+0.09}$ \\
		
		GW190408A & $0.09 ^{+0.13} _{-0.13}$ & $0.04 ^{+0.14} _{-0.13}$ \\
		GW190412A & $-0.03 ^{+0.05} _{-0.05}$ & $-0.05 ^{+0.07} _{-0.09}$ \\
		GW190421A & $-4.04 ^{+15.61} _{-7.53}$ &  --\\
		GW190503A & $1.51 ^{+7.08} _{-1.84}$ & $0.15 ^{+8.25} _{-0.32}$ \\
		GW190512A & $0.32 ^{+0.14} _{-0.47}$ & $-0.13 ^{+1.09} _{-0.21}$ \\
		GW190513A & $0.22 ^{+0.50} _{-0.24}$ & $0.59 ^{+1.96} _{-0.60}$ \\
		GW190517A & $-0.25 ^{+1.36} _{-1.39}$ & $-0.54 ^{+2.46} _{-1.70}$ \\
		GW190519A & $-0.03 ^{+34.39} _{-34.83}$ &  --\\
		GW190521A & $0.04 ^{+4.61} _{-4.72}$ &  --\\
		GW190521B & $-0.03 ^{+0.11} _{-0.09}$ & $-0.00 ^{+0.08} _{-0.08}$ \\
		GW190602A & $0.13 ^{+2.18} _{-2.38}$ & $0.58 ^{+1.77} _{-2.61}$ \\
		GW190630A & $0.08 ^{+0.13} _{-0.11}$ & $0.10 ^{+0.18} _{-0.12}$ \\
		GW190706A & $-0.02 ^{+2.31} _{-2.30}$ &  --\\
		GW190707A & $0.03 ^{+0.07} _{-0.07}$ & $0.02 ^{+0.05} _{-0.07}$ \\
		GW190708A & $0.01 ^{+0.07} _{-0.06}$ & $0.02 ^{+0.07} _{-0.08}$ \\
		GW190720A & $0.03 ^{+0.08} _{-0.06}$ & $0.04 ^{+0.10} _{-0.19}$ \\
		GW190727A & $-0.92 ^{+34.95} _{-33.71}$ & $-0.76 ^{+2.82} _{-1.50}$ \\
		GW190728A & $0.05 ^{+0.09} _{-0.08}$ &  --\\
		GW190814A & $-0.13 ^{+0.17} _{-0.09}$ & $0.01 ^{+0.03} _{-0.19}$ \\
		GW190828A & $-0.04 ^{+0.13} _{-0.10}$ & $-0.05 ^{+0.12} _{-0.13}$ \\
		GW190828B & $0.11 ^{+0.15} _{-0.12}$ & $0.09 ^{+0.17} _{-0.73}$ \\
		GW190910A & $0.00 ^{+1.79} _{-2.25}$ & $-0.02 ^{+0.13} _{-0.16}$ \\
		GW190915A & -- & $-1.90 ^{+2.75} _{-0.58}$ \\
		GW190924A & $-0.02 ^{+0.16} _{-0.11}$ & $0.00 ^{+1.07} _{-1.28}$ \\
    \end{tabular}
    \end{ruledtabular}
    \caption{Constraints on $\delta_2$ for the BBH events in GWTC-1 and GWTC-2 with the IMRPhenomPv2 and SEOBNRv4P waveform models. The reported uncertainties correspond to the 90\% confidence limit. -- means that the data are not available. See the text for more details.}
    \label{tab:delta2}
\end{table}

\begin{figure*}
    \centering
    \includegraphics[width=0.95\textwidth]{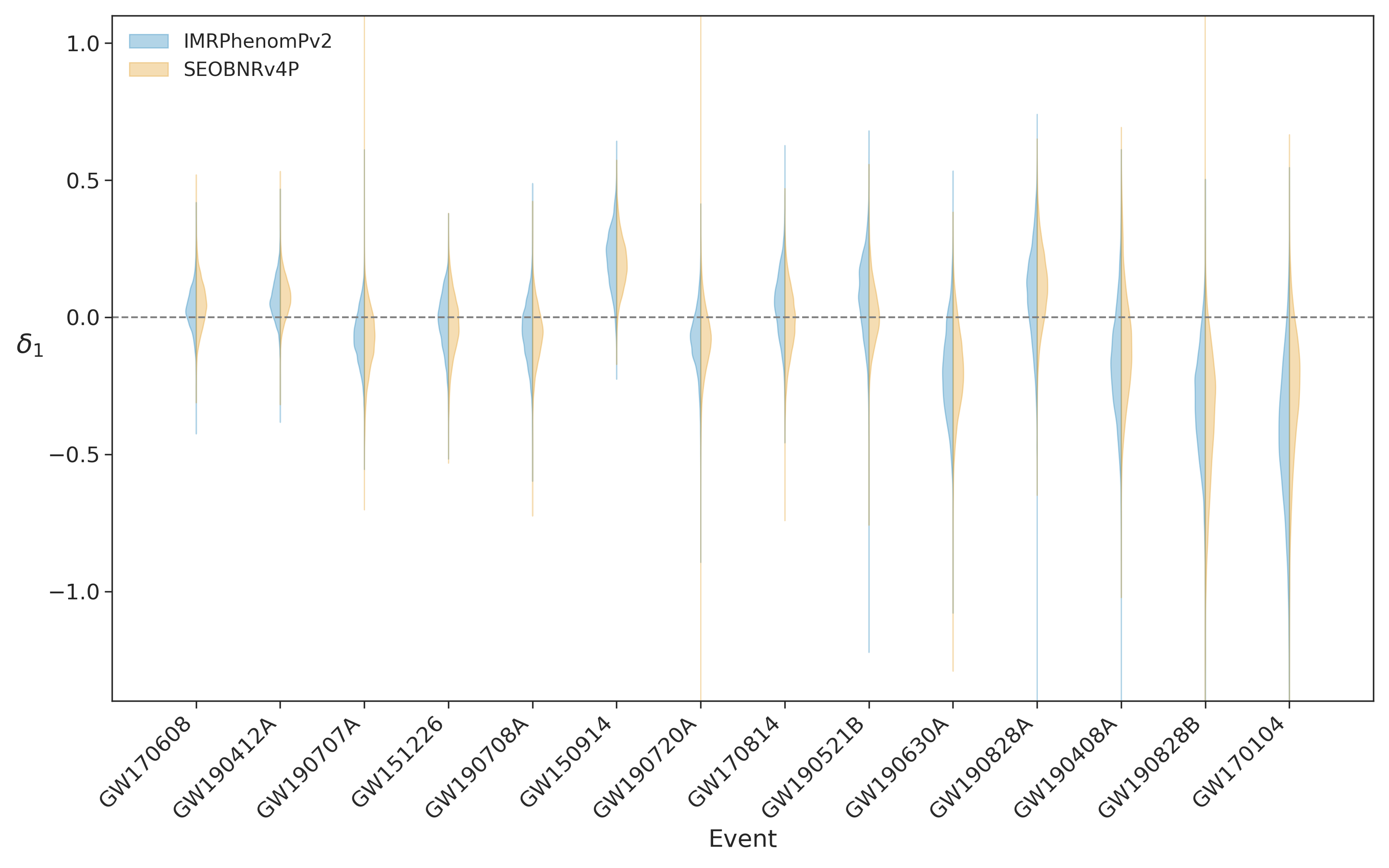}\\
    \vspace{0.3cm}
    \includegraphics[width=0.95\textwidth]{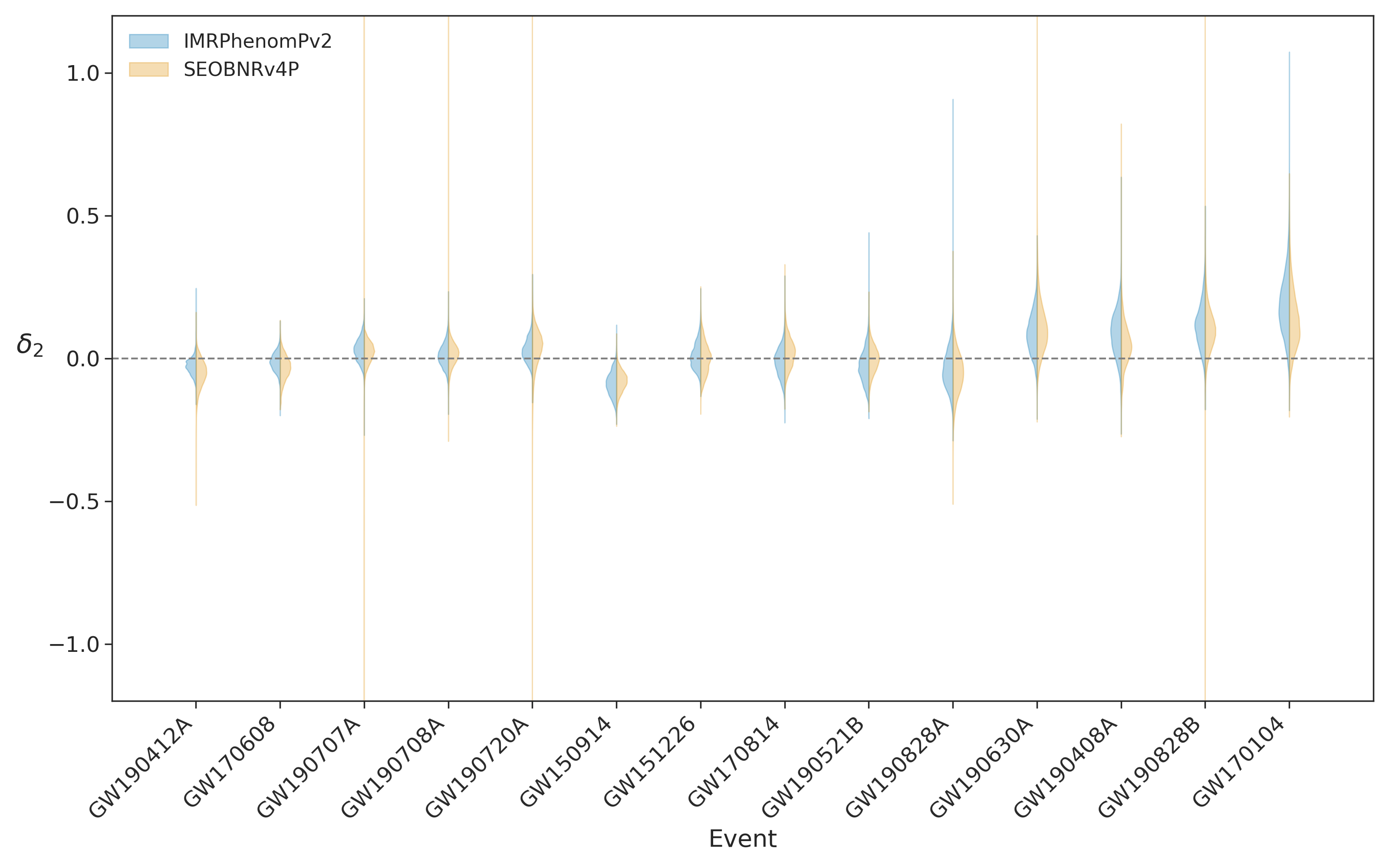}
    \caption{Constraints on $\delta_1$ (top panel) and $\delta_2$ (bottom panel). 
    Here we only show the events for which the 90\% confidence level constraints are smaller than 1, as our analysis assumes $|\delta_1| , |\delta_2| \ll 1$ and considers only the leading-order terms.}
    \label{fig:constraints}
\end{figure*}

Fig.~\ref{fig:constraints} shows the constraints for $\delta_1$ and $\delta_2$ for both the waveform models for the events in which both data are available. We use the 90\% intervals of the IMRPhenomPv2 model as a criterion to decide which events have stronger constraints and the events are arranged in an ascending order. In Tab.~\ref{tab:delta1} and Tab.~\ref{tab:delta2}, we instead list the constraints for all the events in a chronological order. We note that the constraints on  $\delta_1$ and $\delta_2$ are obtained employing Eqs.~\ref{eq:d1-p4} and \ref{eq:d2-p5}. However, these two equations have been derived at leading-order in the deformation parameters $\delta_1$ and $\delta_2$, so they are valid only if $|\delta_1| , |\delta_2| \ll 1$. So in Tab.~\ref{tab:delta1} and Tab.~\ref{tab:delta2} only the constraints $|\delta_1| , |\delta_2| \ll 1$ are consistent with our approximations.

\begin{table}
    \begin{ruledtabular}
    \begin{tabular}{ccc}
        Parameter & IMRPhenomPv2 &  SEOBNRv4P \\
        \hline
        $\delta_1$  & $-0.02\pm0.04$ & $0.02\pm0.05$ \\
		$\delta_2$  & $0.01\pm0.02$ &  $0.00\pm0.02$\\
    \end{tabular}
    \end{ruledtabular}
     \caption{Combined constraints on $\delta_1$ and $\delta_2$ assuming that all sources have the same value of the deformation parameter and that the measurements of $\delta_1$ and $\delta_2$ for every source follow a Gaussian distribution. The reported uncertainties correspond to the 90\% confidence limit. }
    \label{tab:combined}
\end{table}

\section{Discussion and conclusions} \label{sec:conclusion}

In this work, we have tested the Kerr hypothesis within the KRZ framework by using GW data of BBH merger events. Within our approximations, we can constrain the deformation parameters $\delta_1$ and $\delta_2$, while we are not sensitive to the other deformation parameters ($\delta_3$, $\delta_4$, $\delta_5$, and $\delta_6$). From the publicly available data in GWTC-1 and GWTC-2, all events are consistent with the hypothesis that the objects are Kerr BHs. Weak constraints are obtained for the events with low signal-to-noise ratios (SNR) or with high values of the BH mass.

We note that the most stringent constraint on $\delta_1$ is obtained from the event GW170608, which was already in GWTC-1 and was inferred in Ref.~\cite{Cardenas-Avendano_Nampalliwar_Yunes_2020}. The new events reported in GWTC-2 cannot do better, with the constraint from GW190412A only slightly worse than the constraint from GW170608. In GW170608, the masses of the two BHs of the binary system are low, $11.0~M_{\odot}$ and $7.6~M_{\odot}$, and the SNR is quite high, 15.6~\cite{GWTC1}. For the deformation parameter $\delta_2$, the most stringent constraint comes from the event GW190412A, where the BH masses are, respectively, $30.1~M_{\odot}$ and $8.3~M_{\odot}$ and the SNR is 18.9~\cite{GWTC2}.

If we assume that the values of the deformation parameters should be the same for every BH, we can combine the constraints of all events together to increase the SNR and get stronger constraints. Such a possibility depends on the specific gravity theory. In GR, this is trivially true because all objects should be Kerr BHs and the deformation parameters should always vanish. In models violating the no-hair theorem, astrophysical BHs would be specified by other parameters in addition to the mass and the spin angular momentum: in such a case, every source could have potentially different values of the deformation parameters (exactly like they can have different values of $M$ and $a_*$). In models in which the Kerr hypothesis is violated but some form of the no-hair theorem still holds, astrophysical BHs would not be the Kerr BHs of GR but they would be still characterized only by the mass and the spin angular momentum, so the deformation parameters would have the same values for every source. We note that there are also models predicting BHs with ``secondary hairs'' and an example is Einstein-dilaton-Gauss-Bonnet gravity~\cite{Yagi:2016jml}: in such a model, the scalar charge of the BH is not an independent quantity and its value is instead determined by the BH mass. In such a case, all BHs with the same (similar) value of the mass have the same (similar) value of the scalar charge. Tab.~\ref{tab:combined} shows the constraints on $\delta_1$ and $\delta_2$ when we combine all data together. These constraints are derived by multiplying the likelihoods to obtain a combined posterior \cite{Zimmerman:2019wzo} assuming that the measurements of the deformation parameters follow a Gaussian distribution (which is not true but it is an approximation). The reported uncertainties still correspond to the 90\% confidence interval.

We can compare the constraints derived in the present study from GW data with the constraints inferred with electromagnetic techniques. We note that, within the framework and the approximations adopted in the present study, it is relatively straightforward to translate the constraints on $\delta_1$ and $\delta_2$ into the constraints on another deformation parameter: we just need to repeat the calculations in Sec.~\ref{sec:gw_constraint} and find the counterpart of Eqs.~\ref{eq:d1-p4} and \ref{eq:d2-p5} for the new deformation parameter. However, this is not the case for the electromagnetic constraints, which require the construction of the theoretical model with the new deformation parameter and a new analysis of the data with the updated model. For this reason, here we limit our discussion to the comparison of the results presented in this paper and those reported in Paper~I and Paper~II.
We also note that in Refs.~\cite{Volkel:2019muj,Volkel:2020daa} the authors studied the possibility of constraining the deformation parameters of the RZ metric from the ringdown phase of BBH mergers; their analysis was limited to simulated data, so we cannot include their results in our discussion, but it is an alternative and complementary method for testing BHs to bear in mind.

\begin{figure*}
    \centering
    \includegraphics[width=0.49\textwidth]{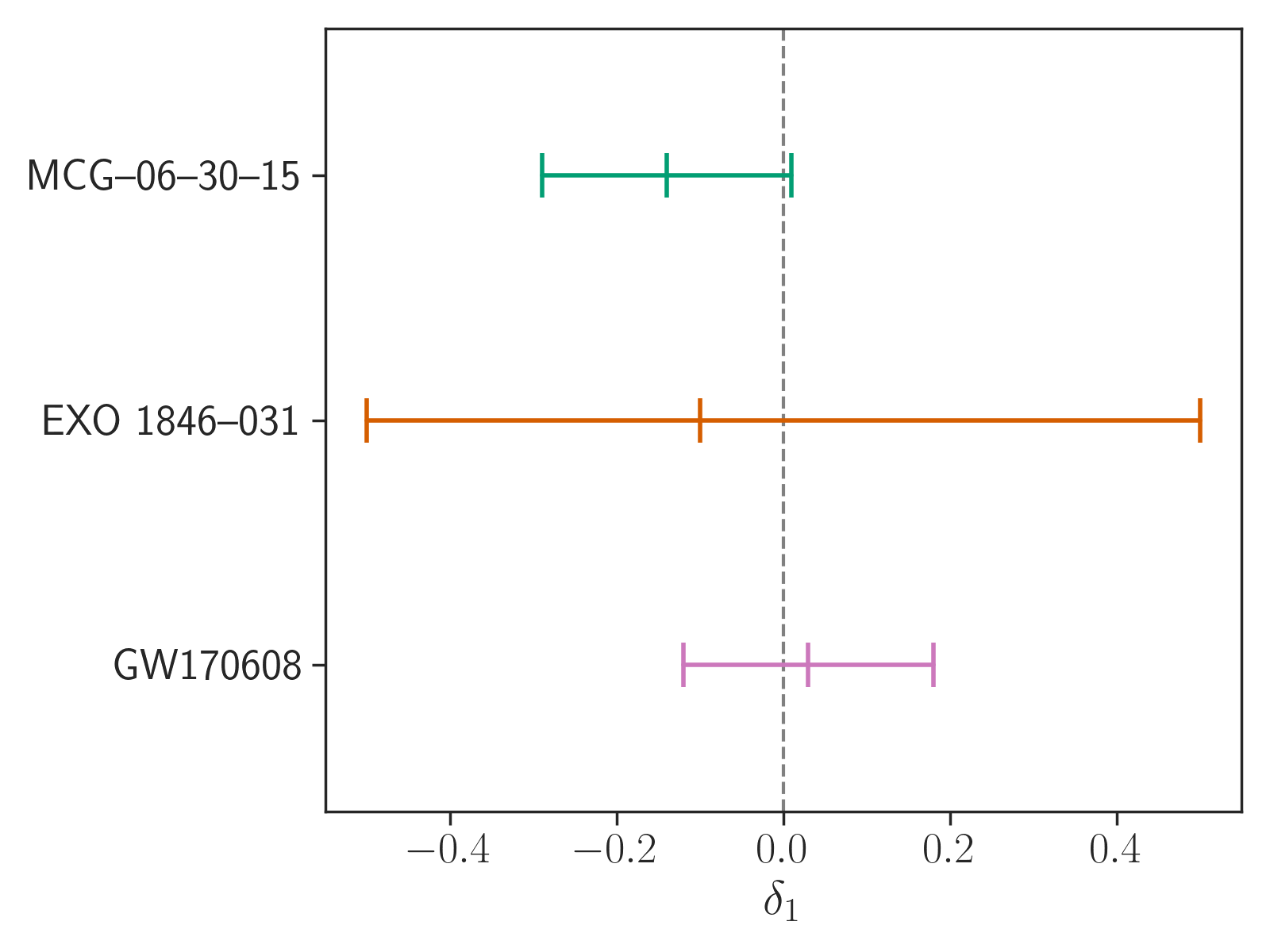}
    \includegraphics[width=0.49\textwidth]{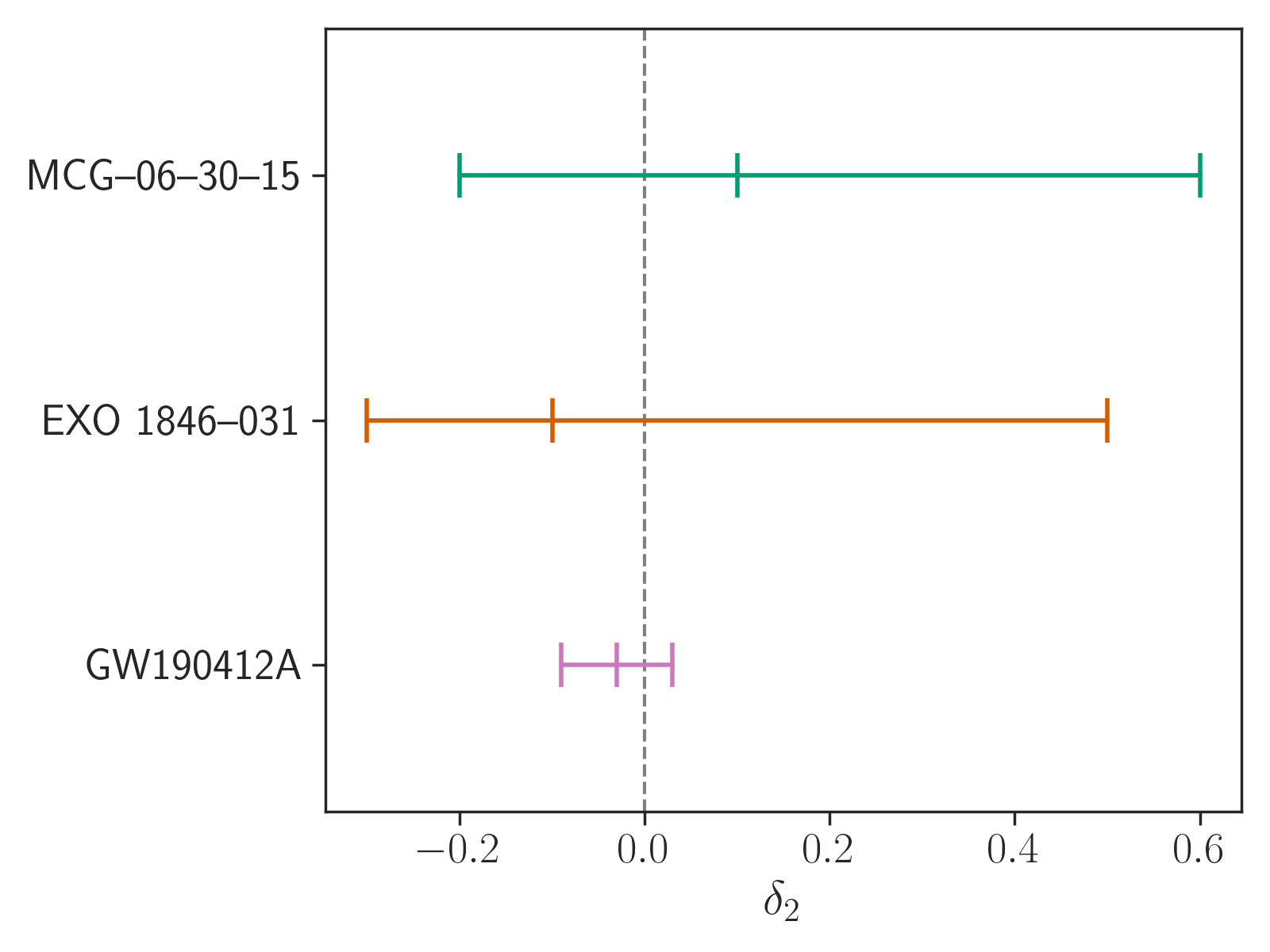}
    \caption{2-$\sigma$ constraints on $\delta_1$ (left panel) and $\delta_2$ (right panel) from the analysis of X-ray observations of MCG--06--30--15 (Paper~I) and EXO~1846--031 (Paper~II) and from the analysis of GW events (this paper, GW170608 for $\delta_1$ and GW190412A for $\delta_2$).}
    \label{fig:xray_v_gw}
\end{figure*}

In Paper~I, we analyzed \textsl{NuSTAR} and \textsl{XMM-Newton} data of the supermassive BH at the center of the galaxy MCG--06--30--15 with \texttt{relxill\_nk} \cite{Bambi:2016sac,Abdikamalov:2019yrr,Abdikamalov:2020oci}, which is a model specifically designed to test the Kerr hypothesis from the analysis of the reflection spectrum of accreting BHs. While X-ray reflection spectroscopy can potentially constrain all deformation parameters of the KRZ metric, we found that the impact of $\delta_3$, $\delta_4$, and $\delta_6$ on the reflection spectrum is too weak, even for the next generation of X-ray satellites. In the end, only the deformation parameters $\delta_1$, $\delta_2$, and $\delta_5$ were constrained. The 2-$\sigma$ constraints on $\delta_1$ and $\delta_2$ of our analysis are
\be
\delta_1 = -0.14 \pm 0.15 \, , \quad \delta_2 = 0.1^{+0.5}_{-0.3} \, .
\ee 
In Paper~II, we analyzed with \texttt{relxill\_nk} a \textsl{NuSTAR} spectrum of the stellar-mass BH in the X-ray binary system EXO~1846--031. The 2-$\sigma$ constraints on $\delta_1$ and $\delta_2$ from this source are
\be
\delta_1 = -0.1^{+0.6}_{-0.4} \, , \quad \delta_2 = -0.1^{+0.6}_{-0.2} \, .
\ee
We note that the constraints from MCG--06--30--15 and EXO~1846--031 are not inferred assuming $|\delta_1| , |\delta_2| \ll 1$ and a vanishing BH spin parameter. The latter is instead determined in the fit together with the values of the deformation parameters. These constraints from X-ray data are clearly weaker than those obtained in the present paper from GW data; see Fig.~\ref{fig:xray_v_gw}, which shows the 2-$\sigma$ constraints on $\delta_1$ and $\delta_2$ from MCG--06--30--15, EXO~1846--031, GW170608 (for $\delta_1$), and GW190412A (for $\delta_2$). However, we note that such a conclusion cannot be generalized for every deformation parameter. For example, X-ray data can constrain the deformation parameter $\alpha_{13}$ of the Johannsen metric~\cite{Johannsen:2013szh} better than GW data, as shown in Ref.~\cite{Tripathi:2021rqs}. This is perfectly understandable: GW and X-ray data test the Kerr hypothesis from the analysis of different relativistic effects, and therefore some deviations from the Kerr solution can be more easily measured with GW tests and other deformations can be better constrained from X-ray data.

Last, we note that the constraints on $\delta_1$ and $\delta_2$ derived in the present paper cannot be immediately translated into constraints of coupling constants of specific non-GR theories, at least for the most popular ones. In general, if we consider a specific gravity theory, the gravitational wave signal will depend on both the black hole metric and the field equations of the theory. The gravitational wave signal of some popular non-GR theories within the ppE formalism was studied in Ref.~\cite{Yunes_Yagi_Pretorius_2016}, and summarized in their Tab.~III. Among the models considered in Ref.~\cite{Yunes_Yagi_Pretorius_2016}, only dynamical Chern-Simons gravity has the leading order correction entering at 2PN, as in the case of our $\delta_1$. However, as already pointed out in Ref.~\cite{Yunes_Yagi_Pretorius_2016}, BBH merger events cannot constrain dynamical Chern-Simons gravity at 2PN because of degeneracies among the BH spins, BH masses, and the coupling constant of the theory. If we consider non-rotating BHs, as in our study, the leading order correction enters at very high PN order. For the other popular non-GR theories, the leading order correction enters at a different PN order with respect to our $\delta_1$ and $\delta_2$ (see, again, Tab.~III in Ref.~\cite{Yunes_Yagi_Pretorius_2016}), and constraining those theories requires an independent analysis.

\begin{acknowledgments}
We thank Alejandro C\'ardenas-Avenda\~no for useful discussions and earlier collaboration on the subject of this paper. 
This work was supported by the Innovation Program of the Shanghai Municipal Education Commission, Grant No.~2019-01-07-00-07-E00035, the National Natural Science Foundation of China (NSFC), Grant No.~11973019, and Fudan University, Grant No.~JIH1512604.
S.S. also acknowledges support from the China Scholarship Council (CSC), Grant No.~2020GXZ016646.
We thank the LVC for making their data public.
\end{acknowledgments}

\appendix

\section{GW constraints on the deformation parameter $\alpha_{13}$ of the Johannsen metric}
\label{appendix:johannsen}

In the past few years, tests of the Kerr hypothesis with BH X-ray data have mainly employed the Johannsen metric and constrained the deformation parameter $\alpha_{13}$ \cite{Johannsen:2013szh}; see, e.g., Refs.~\cite{Bambi:2021chr,Tripathi:2021rqs} for a summary of all available constraints to date. It is thus useful to report here the constraints on $\alpha_{13}$ from the BBH events in GWTC-1 and GWTC-2.

We consider equatorial circular orbits ($\theta = \pi/2$ and $\dot{\theta} = 0$) and a vanishing spin parameter ($a_* = 0$). In such a limit, the line element of the Johannsen metric reads
\be\label{eq:johannsen_metric}
	d s^2 &=& -\frac{\Sigma \Delta}{B^2} ~d t^{2} + \frac{\Sigma}{\Delta} ~d r^{2}
	+ \Sigma \left( d \theta^{2} + d \phi^{2} \right) \, ,
\ee
where
\be\label{eq:johannsen_coeff}
	&& \Sigma = r^2 \, , \quad \Delta = r^2 - 2Mr \, , \quad B = r^2 A \, , \\
	&& A = 1 + \alpha_{13} ~\left(\frac{M}{r}\right)^3 \, .
\ee

From the normalization of the four-velocity $u^{\mu} u_{\mu} = -1$, we write the effective potential
\be
\label{eq:Veff_full_alpha13}
    V_{\mathrm{eff}} &=& -1 + E^2 + \frac{2 M}{r} + \frac{L^2 (2 M-r)}{r^3} \nonumber\\ 
    && + \frac{2 E^2 M^3 \alpha_{13}}{r^3} +\mathcal{O}[{\alpha_{13}}^{2}] \, .
\ee
and we obtain the modified Kepler's law
\be \label{eq:mod_kep_alpha13}
    \Omega^2 = \frac{M}{r^3} \left[ 1 + \frac{3 M}{r} + \frac{9 M^2}{r^2} - \frac{3 M^2 \alpha_{13}}{r^2} + \mathcal{O} \left({\alpha_{13}}^{2}, \frac{M^3}{r^3} \right) \right] \, . \nonumber\\
\ee
The leading order correction in $\alpha_{13}$ is at 2PN, as it was in the case of $\delta_1$. The formula connecting the deformation parameter $\alpha_{13}$ to the 2PN phase $\varphi_4$ and the 2PN non-GR parameter $\delta \varphi_4$ is
\be
   \alpha_{13} = -\frac{1}{100} \varphi_{4} \delta \varphi_{4} \, .
\ee

\begin{table}
	\begin{ruledtabular}
	\begin{tabular}{ccc}
		Event & $\alpha_{13}$ (IMRPhenomPv2) & $\alpha_{13}$ (SEOBNRv4P) \\
		\hline
		GW150914 & $0.88 ^{+0.72} _{-0.78}$ & $0.78 ^{+0.67} _{-0.61}$  \\
		GW151226 & $-0.03 ^{+0.59} _{-0.72}$ & $-0.11 ^{+0.73} _{-0.80}$  \\
		GW170104 & $-1.71 ^{+1.63} _{-2.10}$ & $-1.00 ^{+1.32} _{-1.79}$  \\
		GW170608 & $0.12 ^{+0.49} _{-0.48}$ & $0.19 ^{+0.63} _{-0.58}$  \\
		GW170814 & $0.21 ^{+0.75} _{-0.79}$ & $-0.06 ^{+0.79} _{-0.88}$  \\
		GW190408A & $-0.69 ^{+1.26} _{-1.24}$ & $-0.45 ^{+1.58} _{-1.24}$  \\
		GW190412A & $0.25 ^{+0.54} _{-0.46}$ & $0.30 ^{+0.51} _{-0.48}$  \\
		GW190421A & $29.74 ^{+14.11} _{-65.03}$ & --  \\
		GW190503A & $-1.69 ^{+4.50} _{-27.30}$ & $-0.93 ^{+2.51} _{-25.83}$  \\
		GW190512A & $-2.95 ^{+4.27} _{-1.24}$ & $0.87 ^{+0.92} _{-5.08}$  \\
		GW190513A & $-2.05 ^{+2.28} _{-3.74}$ & $-2.81 ^{+3.05} _{-2.65}$  \\
		GW190517A & $1.57 ^{+6.20} _{-4.85}$ & $1.99 ^{+3.01} _{-2.32}$  \\
		GW190519A & $-4.74 ^{+124.28} _{-116.46}$ & --  \\
		GW190521A & $-0.13 ^{+15.37} _{-15.60}$ & --  \\
		GW190521B & $0.32 ^{+0.86} _{-0.99}$ & $0.03 ^{+0.80} _{-0.69}$  \\
		GW190602A & $-0.68 ^{+8.54} _{-7.39}$ & $-1.63 ^{+9.09} _{-6.57}$  \\
		GW190630A & $-0.80 ^{+1.13} _{-1.08}$ & $-0.77 ^{+1.02} _{-1.05}$  \\
		GW190706A & $0.40 ^{+7.72} _{-8.31}$ & --  \\
		GW190707A & $-0.27 ^{+0.62} _{-0.64}$ & $-0.30 ^{+0.70} _{-0.88}$  \\
		GW190708A & $-0.19 ^{+0.68} _{-0.77}$ & $-0.24 ^{+0.65} _{-0.72}$  \\
		GW190720A & $-0.32 ^{+0.69} _{-0.85}$ & $-0.31 ^{+0.81} _{-0.82}$  \\
		GW190727A & $14.24 ^{+109.18} _{-138.96}$ & $5.43 ^{+2.79} _{-12.77}$  \\
		GW190728A & $-0.55 ^{+0.84} _{-1.07}$ & --  \\
		GW190814A & $-0.05 ^{+2.39} _{-0.31}$ & --  \\
		GW190828A & $0.36 ^{+1.07} _{-1.23}$ & $0.45 ^{+0.94} _{-0.92}$  \\
		GW190828B & $-1.24 ^{+1.27} _{-1.37}$ & $-1.41 ^{+1.42} _{-2.07}$  \\
		GW190910A & $-0.00 ^{+9.43} _{-2.10}$ & $0.12 ^{+1.24} _{-1.31}$  \\
		GW190915A & -- & $-0.31 ^{+8.72} _{-8.19}$  \\
		GW190924A & $-0.34 ^{+0.97} _{-1.48}$ & $-0.72 ^{+1.59} _{-2.42}$  \\
		\hline
		Combined & $-0.10 \pm 0.16$ & $0.08 \pm 0.19$  \\
	\end{tabular}
	\end{ruledtabular}
	\caption{Constraints on $\alpha_{13}$ for the BBH events in GWTC-1 and GWTC-2 with the IMRPhenomPv2 and SEOBNRv4P waveform models. The reported uncertainties correspond to the 90\% confidence limit. -- means that the data are not available. See the text for more details.}
	\label{tab:alpha13}
\end{table}

Tab.~\ref{tab:alpha13} shows the constraints on $\alpha_{13}$ for the BBH events in GWTC-1 and GWTC-2 with the IMRPhenomPv2 and SEOBNRv4P waveform models. Only the constraints with $|\alpha_{13}| \ll 1$ are consistent with our approximations. At the end of the table, we report the combined constraint from all events under the assumption that the deformation parameter follows a Gaussian distribution. The events GW170608 and GW190412A provide the strongest constraints among all single-event constraints. For the deformation parameter $\alpha_{13}$, BH X-ray data provide more stringent constraints \cite{Bambi:2021chr,Tripathi:2021rqs} than GW data.
The constraints on $\alpha_{13}$ from the shadow of the M87 BH are reported in Ref.~\cite{EventHorizonTelescope:2020qrl}.

\nocite{*}

\bibliographystyle{apsrev4-1}
\bibliography{references}

\end{document}